\documentclass[sigconf]{acmart}

\usepackage{booktabs} 

\setcopyright{acmlicensed}
\acmConference[DAPA '19]{DAPA 2019 WSDM Workshop on Deep matching in Practical Applications}{February 15th, 2019}{Melbourne, Australia} 
\acmYear{2019}
\copyrightyear{2019}
\acmDOI{}
\acmISBN{}
\acmPrice{}

\begin{document}
\title{Learning From Weights: A Cost-Sensitive Approach For Ad Retrieval}

\author{Nikit Begwani\textsuperscript{*}}
\affiliation{
 \institution{Microsoft India}
}
\email{nibegwan@microsoft.com}

\thanks{* These authors contributed equally to this work.}

\author{Shrutendra Harsola\textsuperscript{*}}
\affiliation{
	\institution{Microsoft India}
}
\email{shharsol@microsoft.com}

\author{Rahul Agrawal}
\affiliation{
	\institution{Microsoft India}
}
\email{rahulagr@microsoft.com}

\begin{abstract}

Retrieval models such as CLSM is trained on click-through data which treats each clicked query-document pair as equivalent. While training on click-through data is reasonable, this paper argues that it is sub-optimal because of its noisy and long-tail nature (especially for sponsored search). In this paper, we discuss the impact of incorporating or disregarding the long tail pairs in the training set. Also, we propose a weighing based strategy using which we can learn semantic representations for tail pairs without compromising the quality of retrieval. We conducted our experiments on Bing sponsored search and also on Amazon product recommendation to demonstrate that the methodology is domain agnostic.

Online A/B testing on live search engine traffic showed improvements in clicks (11.8\% higher CTR) and as well as improvement in quality (8.2\% lower bounce rate) when compared to the unweighted model. We also conduct the experiment on Amazon Product Recommendation data where we see slight improvements in NDCG Scores calculated by retrieving among co-purchased product.
\end{abstract}

\keywords{Neural Networks, Cost-sensitive Learning, Sponsored Ads}

\maketitle

\vspace{-2mm}
\section{Introduction}

\textbf{Objective} : This paper formulates the problem of learning neural semantic models for IR as a cost-sensitive learning problem. It explores various costing (weighing) techniques for improving neural semantic models, specifically the CLSM model \cite{CLSM}. In online retrieval, we have millions of documents with which query similarity needs to be calculated within milliseconds and thus query-document word level interaction is not possible and hence we rely on representation based model and CLSM is the state-of-the-art representation based semantic model.

\textbf{Sponsored Ad Retrieval} : Search engines generate most of their revenue via sponsored advertisements, which are displayed alongside organic search results. Ad retrieval system is designed to select ads in response to the user queries. Historically, advertisers created their campaign by providing ad and a set of queries (bid terms) that they want to display their ad on. This scenario is called an exact match. But it is not possible for advertisers to provide bid terms to cover all tail queries by exact match. An advanced match is used to overcome this issue, where user queries are semantically matched with ads.  
Each serving of an ad, in response to a user query, is called an impression. For a query-ad pair, total clicks divided by total impressions over a time interval is defined as click-through rate (CTR), while the percentage of times a user returns back immediately after clicking an ad is referred as bounce rate (BR). High click-through rate and low bounce rate is desirable for a sponsored search system.

Earlier information retrieval techniques matched queries with ads based on syntactic similarity \cite{BM25}. 
Lot of recent research went in developing semantic retrieval techniques like LSA \cite{LSA}, pLSI \cite{PLSI} and LDA \cite{LDA}, which maps query and document in lower dimensional semantic space, where match can happen even with no token overlap.

\begin{figure}
	\includegraphics[width=0.8\columnwidth]{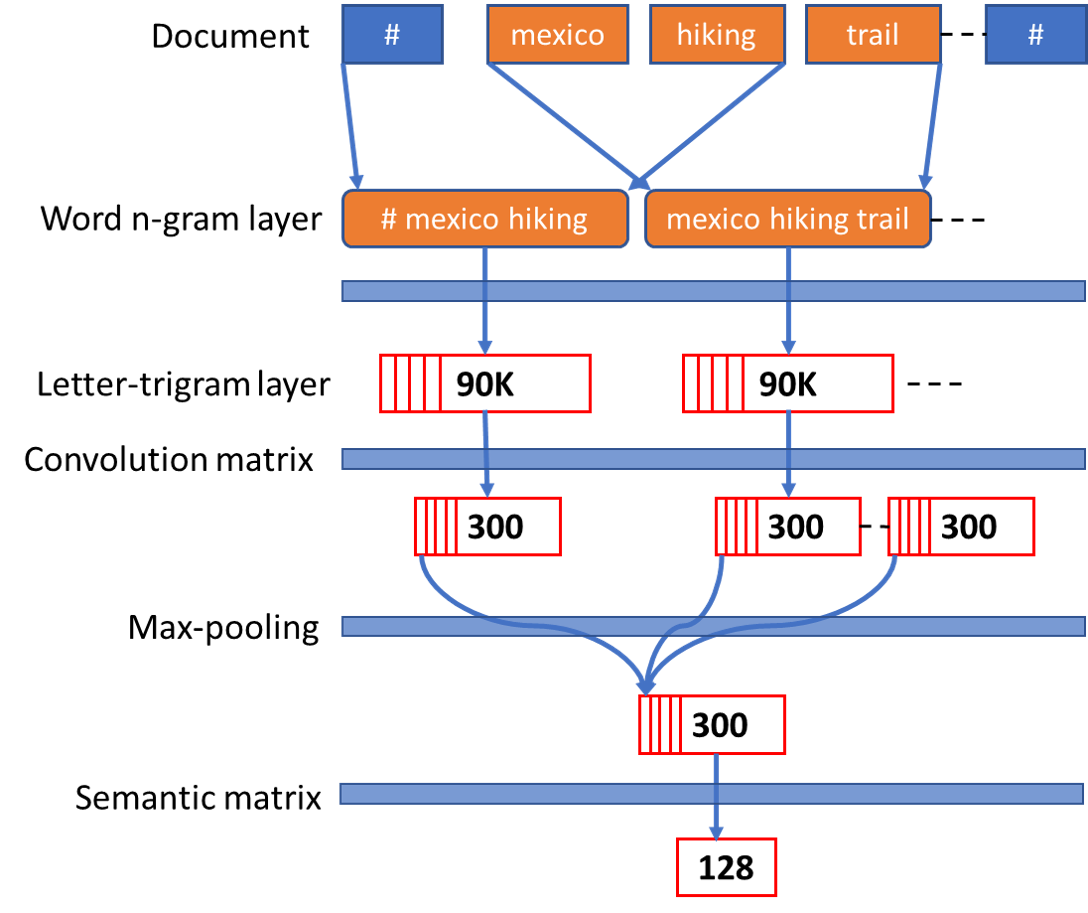}
	\caption{CLSM Architecture}
	\label{fig:CLSMArchitecture}
		\vspace{-6mm}
\end{figure}

\textbf{DSSM} : Recently, there has been a shift towards neural network based semantic models trained using click-through data. DSSM \cite{DSSM} is one such representation based neural network model. It takes a bag of words representation of historically clicked query-document pairs and maps them in lower dimension space using a discriminative approach. It uses a set of non-linear layers to generate query and document semantic vectors. The learning objective is to maximize the cosine similarity between the vectors of clicked query-document pairs. 

\textbf{CLSM} : Bag-of-words representation for query/document is used in DSSM, which is not suitable for capturing contextual structures. CLSM \cite{CLSM} tries to solve this issue by running a contextual sliding window over the input word sequence. As shown in figure \ref{fig:CLSMArchitecture}, CLSM has letter trigram based word hashing layer, followed by convolution layer based sliding window which generates a local contextual feature vector for each word within its context window. These local features are then aggregated using a max-pool layer to generate the global feature vector, which is then fed to a fully connected layer to generate the high-level semantic vector for query/document.

\begin{figure}
	\includegraphics[width=0.8\columnwidth]{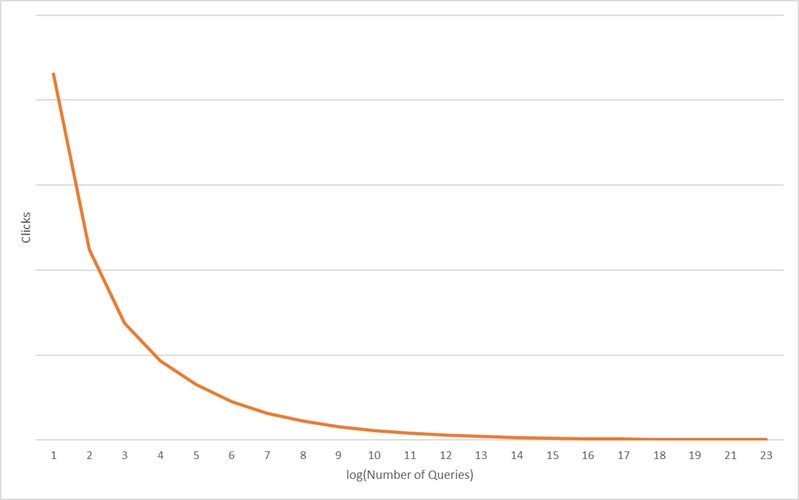}
	\caption{Clicks v/s number of queries showing the long tail nature of click-through data. Note that x-axis is on log-scale.}
	\label{fig:ClickQueries}
	\vspace{-5mm}
\end{figure}

\textbf{Motivation} : Current semantic models trained on click-through data treat all historical clicked query-document pairs as equally important, which is not true. For example, an ad related to "thyroid treatment" can receive a click from a wide variety of queries like Hashimoto disease, hypothyroidism, medication hypothyroidism, Hashimoto, swollen glands neck, heart conditions, liver diseases, Perthes diseases etc. Some of these queries are very specific, while others are generic. Treating all these pairs as equal while training can result in model learning only generic patterns. Other dimensions of the problem are the noise in click-through data (i.e. not all clicked ads for a query are relevant to the query) and long tail nature of click-through data (see figure \ref{fig:ClickQueries}). For example, we can fairly confidently say that a query-ad pair having 95 clicks from 100 impressions is relevant, but not much can be said about a query-ad pair with 1 click from 1 impression.
One solution is to generate training data by applying minimum impression, click and CTR thresholds on query-ad pairs. But this can result in the filtering of most of the tail queries (since long tail queries never repeat, so all query-ad pairs for such queries will have only one impression). This can result in below par performance of semantic models for tail queries. This is not an acceptable solution since semantic models are mainly used for an advanced match in tail queries.

To address this issue, we propose that all clicked query-ad pairs should be used for training semantic models. Further, we propose different costing (weighing) techniques on click data, so that model learns from important pairs.

\textbf{Contributions} : This paper formulates the neural semantic model training as a cost-sensitive learning problem. It propose approaches for re-weighing training data and guiding principles for the same.
Online A/B testing of the weighted model on live traffic shows 11.8\% gains in clicks and CTR and 8.2\% improvement in terms of quality measured using bounce rate(BR) as compared to the unweighted model. Further evaluation of weighted-CLSM model on amazon co-purchased dataset shows 0.27 absolute increase in NDCG@1 and 0.25 absolute increase in NDCG@3 over the unweighted model.

\section{Related Work}

\subsection{Traditional IR Techniques}
Many IR techniques have been proposed for modeling contextual information between queries and documents \cite{BM25} \cite{CLSMRef2} \cite{CLSMRef11} \cite{CLSMRef12} \cite{CLSMRef22} \cite{CLSMRef24} \cite{CLSMRef25} \cite{CLSMRef26}. Classical TF-IDF and BM25 (\citet{BM25}) based techniques are based on a bag of words representation. These approaches are further extended by modeling term/n-gram dependencies using Markov Random Field (\citet{CLSMRef25}), Latent Concept Expansion (\citet{CLSMRef26}), dependence model (\citet{CLSMRef11}) and phrase translation model (\citet{CLSMRef12}).
\vspace{-1mm}
\subsection{Semantic Models for IR}
Classical IR techniques based on lexical matching can fail to retrieve relevant documents due to language/vocabulary mismatch between query and documents. Latent Semantic Models aim to solve this issue by mapping both query and document into a lower dimensional semantic space and then, matching the query with documents based on vector similarity in the latent space. Techniques in this area include Latent Semantic Analysis (\citet{LSA}), Probabilistic latent semantic indexing (\citet{PLSI}), Latent Dirichlet Allocation (\citet{LDA}), LDA based document models (\citet{LDA-Retrieval}), Bi-Lingual Topic Model (\citet{BLTM}) etc.

\textbf{Representation based neural models:} These models use a neural network to map both query and document to low dimensional latent embedding space and then perform matching in latent space. Models proposed in this category are DSSM (\citet{DSSM}), CLSM (\citet{CLSM}), ARC-I (\citet{ARC-I}) etc.

\textbf{Interaction based neural models:} These models compute interaction (syntactic/semantic similarity) between each term of query and document. These interactions are further summarized to generate a matching score. Multiple models have been proposed in this category such as DRMM (\citet{DRMM}), MatchPyramid (\citet{MatchPyramid}), aNMM (\citet{aNMM}), Match-SRNN (\citet{Match-SRNN}), K-NRM (\citet{K-NRM}) etc.

\vspace{-1mm}
\subsection{Learning to rank}
Learning to rank (LTR) models for IR aim to learn a ranking model which can rank documents for a given query. \citet{LTR} categorized LTR approaches into three categories based on learning objective: 
Pointwise approaches (\citet{RegLoss1}, \citet{RegLoss2}, \citet{ClassLoss1}), Pairwise approaches (RankNet \cite{RankNet}) and Listwise approaches (LambdaRank \cite{LambdaRank}, ListNet \cite{ListNet}, ListMLE \cite{ListMLE}). 

\vspace{-1mm}
\subsection{Cost Sensitive Learning}
Cost-sensitive learning refers to the problem of learning models when different misclassification errors incur different penalties (\citet{CostSensitive:Elkan}). It was shown by \citet{CostSensitive:John} that learning algorithms can be converted into cost-sensitive learning algorithms by cost-proportionate weighing of training examples or by rejection based subsampling. \citet{CostSensitive:WeightedML} showed that, for the cost-sensitive scenario, the empirical loss is upper bounded by negative weighted log likelihood. It further argues that weighted maximum likelihood should be the standard approach for solving cost-sensitive problems.

\section{Cost-sensitive Semantic Model}

\subsection{Proposed Formulation}
Neural semantic models like CLSM \cite{CLSM} learns using click-through data. For a given query, it models the posterior probability of positive/clicked doc ($D^+$) as softmax over positive doc and J randomly selected unclicked documents.
\begin{equation*}
    P(D^+/Q) = \frac {exp(R(Q,D^+))} {\sum\limits_{D` \in \textbf{D}} exp(R(Q,D`))} 
\end{equation*}
where, \textbf{D} contains $D^+$ (clicked doc) and J randomly selected unclicked documents. $R(Q,d)$ represents the cosine similarity between query Q and document d semantic vectors generated by the model.

Model is learned by minimizing the negative log-likelihood of clicked query-ad pairs:
\begin{equation*}
	L(\Lambda) = \sum_{Q \in \textbf{Q}} \;\; \sum_{D^+ \in Clicked(Q)} (-log \, P(D^+/Q))
\end{equation*}
where $\Lambda$ are model parameters, \textbf{Q} is the set of all queries and $Clicked(Q)$ is the set of all documents which have received click when displayed for query Q (based on click logs).

For case of J=1 (i.e. one negatively sampled doc $D^-$ per clicked doc $D^+$), we have:
\begin{equation*}
 \ell = - log \, P(D^+/Q) = -log \frac {exp(R(Q,D^+))} {exp(R(Q,D^+)) + exp(R(Q,D^-))} 
\end{equation*}

\begin{equation*}
    \;\;\;\;\; = log \, (1 + exp(s^- - s^+))
\end{equation*}
\begin{equation*}
    where \;\; s^+ = R(Q,D^+) \;\; and \;\; s^- = R(Q,D^-)     
\end{equation*}
\begin{equation*}
\frac{\partial \ell} {\partial \Lambda} = \frac {exp(s^- - s^+)} {1 + exp(s^- - s^+)} ( \frac{\partial s^-}{\partial \Lambda} - \frac{\partial s^+}{\partial \Lambda} )
\end{equation*}

Assuming true label to be 1 for $D^+$ and 0 for $D^-$, this loss is the same as pair-wise loss from \citet{RankNet}. As discussed earlier, treating all clicked query-document pairs as equally relevant is not optimal since click-through data is noisy and has long tail nature. To address this issue, we propose to assign label $y(Q,D)$ of document D based on its probability of generating a click for query Q. Based on click logs, the probability of click for a query-doc pair can be estimated by its click-through rate. Given these real-valued labels, we can now directly optimize list-wise loss (like DCG). As shown in \citet{LambdaRank}, DCG can be efficiently optimized by modifying gradients as follows:

\begin{figure}
	\includegraphics[width=0.8\columnwidth]{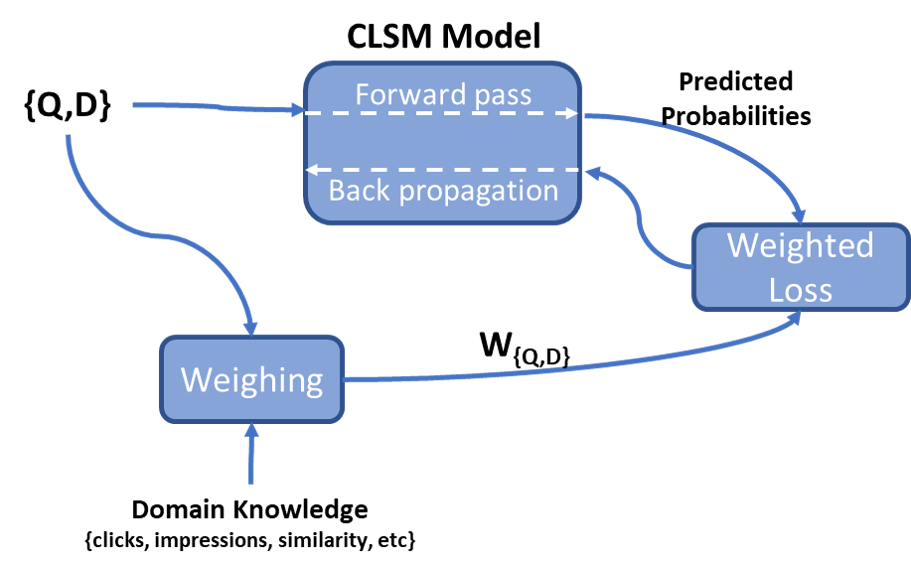}
	\caption{Weighted training of CLSM model incorporates domain knowledge to weigh training pairs.}
	\label{fig:WCDSSM}
\end{figure}

\begin{equation*}
\frac{\partial \ell'} {\partial \Lambda} = \frac {\mid \Delta_{DCG} \mid * \; exp(s^- - s^+)} {1 + exp(s^- - s^+)} ( \frac{\partial s^-}{\partial \Lambda} - \frac{\partial s^+}{\partial \Lambda} )
\end{equation*}

where $\Delta_{DCG}$ is the change in DCG on swapping ranks of $D^+$ and $D^-$.

For a query Q, let $\{D_1,...,D_k\}$ be the top k predicted documents based on model scores with corresponding true labels $\{y(Q,D_1)$,... ,$y(Q,D_k)\}$. Then, DCG@k is defined as:
\begin{equation*}
    DCG@k = \sum_{i=1}^k \frac{y(Q,D_i)}{log_2(1+i)}
\end{equation*}

For training pair $\{Q, D^+=D_j\}$, change in DCG $(\Delta_{DCG})$ on swapping rank position with a negative doc $D^-$ :

(Note that $y(Q,D^-)$ will be 0, since is has zero clicks).
\begin{equation*}
    \mid \Delta_{DCG} \mid = \frac{y(Q,D^+)}{log_2(1+j)}
\end{equation*}

For j=1, i.e. $D^+=D_1$:
\begin{equation*}
	\mid \Delta_{DCG} \mid = {y(Q,D^+)} \;\; 
\end{equation*}

This is equivalent to optimizing following loss function:
\begin{equation*}
	L'(\Lambda) = \sum_{Q \in \textbf{Q}} \;\; \sum_{D^+ \in Clicked(Q)} - log \, P(D^+/Q)^{y(Q,D^+)}
\end{equation*}

\begin{equation} \label{eq:WeightedLoss}
	L'(\Lambda)  = \sum_{Q \in \textbf{Q}} \;\; \sum_{D^+ \in Clicked(Q)} - y(Q,D^+) * log \, P(D^+/Q)
\end{equation}

This can be interpreted as weighing each training point $\{Q,D^+\}$ by weight $y(Q,D^+)$. This shows that in the CLSM scenario, train data weighing is same as optimizing DCG, rather than a pair-wise loss. Figure \ref{fig:WCDSSM} shows a graphical representation of our approach, where the loss for each training point is weighed based on domain knowledge.

The proposed loss function is general in two ways. First, it can be used to learn any representation and interaction based neural semantic model. Second, different weighing strategies (other than CTR) can also be used based on the problem domain.


\subsection{Relation to Cost-Sensitive Learning}
While learning semantic models for Ad retrieval, the cost of misclassification is not the same for all documents. Most sponsored ad systems use cost per click model and hence, try to maximize click-through rates (CTR). So, for a given query, the cost of misclassification is more for a doc with larger expected CTR as compared to a doc with lower expected CTR. With this motivation, we treat the problem of learning semantic models using click data as a cost-sensitive learning problem. As shown in \citet{CostSensitive:John} and \citet{CostSensitive:WeightedML}, the standard approach to solving such problems is to optimize weighted log-likelihood of data, where weights are set according to "costliness" of misclassifying that example.
\begin{equation}
	L'(\Lambda)  = \sum_{Q \in \textbf{Q}} \;\; \sum_{D^+ \in Clicked(Q)} - \; C(Q,D^+) * log \; P(D^+/Q)
\end{equation}
where, $C(Q,D^+)$ is the cost and can be set as 
\begin{equation*}
    C(Q,D^+) = y(Q,D^+) - y(Q,D^-) = y(Q,D^+)
\end{equation*}
(since $y(Q,D^-) = 0$)

This shows that proposed weighted loss function of eq. \ref{eq:WeightedLoss} can also be derived by treating semantic model learning as a cost-sensitive learning problem.


\subsection{Bounds on Ranking Measure}
\citet{LTR:LossFunctions} showed that, for a given query, pair-wise loss upper bounds (1-NDCG).
\begin{equation*}
	1 - NDCG(f;\textbf{x},\mathcal{L}) \leq \frac {\beta_1(s)} {N_n} L^P(f;\textbf{x},\mathcal{L})
\end{equation*}
where $\beta_1(s) = G(K-1)D(1)$, f is the learned ranking function, G is an increasing function (called Gain function), D is a decreasing function (called position discount function), $N_n$ is the DCG of ideal ranking, $\textbf{x}=\{x_1,...,x_n\}$ is the set of documents to be ranked, $\mathcal{L} = \{l(1),...,l(n)\}$  are labels for \textbf{x} and $L^P$ is the pair-wise loss. 
As shown in \citet{LTR:LossFunctions}, this bound can be tightened by introducing weights $W(s)$ proportional to $\beta_1(s)$ in the pair-wise loss as follows:
\begin{equation*}
	\tilde{L}^P(f;\textbf{x},\mathcal{L}) = \sum_{s=1}^{n-1} W(s) \sum_{i=1,l(i)<l(s)}^{n} \phi ( f(x_s) - f(x_i) )
\end{equation*}
where $\phi$ is the logistic function ($\phi(z) = log(1+\exp(-z))$).

Our formulation of eq. \ref{eq:WeightedLoss} can be derived by setting weights as follows:
\begin{equation*}
	W(s) = G(s)D(1) = y(s)
\end{equation*}
Since $D(i) = \frac{1}{log_2(1+i)}$ i.e. $D(1) = 1$. This establishes that proposed CTR based weighting of pair-wise loss tightens the upper bound on (1-NDCG).


\subsection{Weighing Strategies}
\label{sec:WeighingStrategies}
We propose a set of guiding principles, which can used for coming up with weighing strategies:
\begin{enumerate}
	\item Weight should be global in nature i.e. weight for a (query,doc) should be comparable across documents and queries.
	\item Weight should not be biased towards head/torso/tail queries.
	\item Weight should not be biased towards head/torso/tail documents.
	\item Weight should be proportional to the clicks..
\end{enumerate}

(2) and (3) ensures that learned model is not biased towards a particular set of queries or documents, (1) ensures that global threshold tuning can be done on model outputs and (4) ensures that more clickable pairs are ranked higher.

Given click logs over N queries and M documents, let $I_{ij}$, $c_{ij}$ and $w_{ij}$ represent number of impressions, number of clicks and weight for \{$Q_i$, $D_j$\} pair. We can then define following weighing strategies:

 \textbf{nClicks:} Computed as number of clicks for a query-document pair normalized by total clicks for the query over all documents.
	\begin{equation*}
		w_{ij} = { c_{ij} } \big/  {\sum\limits_{m=1}^M  c_{im}}
	\end{equation*}
\textbf{CTR:} Refers to Click through rate and is computed as number of clicks divided by number of impressions.
	\begin{equation*}
		w_{ij} =  { c_{ij} } \big/  { I_{ij}}
	\end{equation*}

\begin{table}
	\begin{tabular}{ c  c  c  c  c  c  }
		\hline
		Weighing Strategy	& P-1 & P-2 & P-3& P-4  \\  \hline
		nClicks		& N & Y & N & Y  \\ 				
		CTR			& Y & Y & Y & Y  \\ \hline
	\end{tabular}
	\caption{Weighing strategies with satisfying principles.}
	\label{tab:WeighingStrategies}
	\vspace{-8mm}
\end{table}

Table \ref{tab:WeighingStrategies} shows the guiding principles satisfied by each of these weighing strategies. CTR satisfy all 4 principles.

\section{Experiments}
In this section, we discuss the experimental results to demonstrate the impact of weighing the click-through data on CLSM model training. We compare following approaches of training CLSM model:
\begin{enumerate}
	\item \textbf{Curated Training:} Only high confidence clicked query-ad pairs are used for training. Where high confidence query-ad pairs are those with CTR greater than the market average.
	\item \textbf{Unweighted Training:} All historically clicked query-ad pairs are used for training with equal weight.
	\item \textbf{Weighted Training:} All historically clicked query-ad pairs are used for training and weights for these pairs are set based on weighting strategies discussed in section \ref{sec:WeighingStrategies}.
\end{enumerate}

These models are compared using two sets of experiments. Firstly, models are evaluated against a human-labeled set of query-ad pairs and AUC numbers are reported to show that weighing doesn't deteriorate the offline metric. Secondly, A/B testing is performed on live search engine traffic and user interaction metrics are reported. 

To prove the efficacy of weighing in a general scenario, we also evaluate the performance of weighted model on Amazon co-purchased dataset discussed in \cite{AmazonDataset} and report nDCG scores.

\subsection{Offline Experiments on Sponsored Search}
\subsubsection{Dataset and Evaluation Methodology}

We take training and evaluation set from Bing sponsored search data for travel vertical. The training data had 11M clicked query-ad pairs and evaluation data set had 154K human-labeled pairs. The ad is represented by ad title (as suggested in \cite{CLSM}). All pairs are then preprocessed such that the text is white-space tokenized, lowercased and alphanumeric in nature, we don't perform any stemming/inflection. 

\begin{table}
	(a) \hspace{1mm}
	\begin{tabular}{  c  c  }
		\hline
		Type of Query & Percentage Share \\  \hline
		Head & 11.25\%  \\ 
		Torso & 35.08\%  \\  
		Tail & 53.67\%  \\  \hline
	\end{tabular}

\vspace{3mm}
(b) \hspace{1mm}
\begin{tabular}{  c  c  c   }		
	\hline
	Total & Positive(1) & Negative(0)  \\ \hline
	154K & 115K & 39K  \\  \hline
\end{tabular}
	\caption{Human-labeled Query-Ad evaluation set}	
	\label{tab:EvalData}
		\vspace{-9mm}
\end{table}
Evaluation data collected from search log is labeled by human judges into positive(1) and negative(0) pairs and Table \ref{tab:EvalData} shows the distribution. Performance of different models is evaluated using Area Under the ROC Curve (AUC-ROC) and Area Under the Precision-Recall Curve (AUC-PR).  Note that, the candidate set of documents for a query changes very frequently in sponsored search and labeling of complete candidate document set for queries is not feasible. Hence, a random set of query-document pairs were labeled and AUC is reported.

\subsubsection{Model Comparisons}
First set of models (Curated and Unweighted) treat all training pairs to be of equal importance i.e. no weighing (Row 1 and Row 2 in Table \ref{tab:TravelResults}). Further, we create a second set of weighted models where each query-ad pair is weighted using strategies discussed in section \ref{sec:WeighingStrategies} i.e. nClicks and CTR. 

The neural network weights for each of the model were randomly initialized as suggested in \cite{ExpRef}. The models are trained using mini-batch based stochastic gradient descent with the mini-batch size of 1024 training samples. Each training point was associated with 4 randomly selected negative samples during training.
\begin{table}	
	\begin{tabular}{c  c  c  c   }		
		\hline
		\# 	& Model Type		& AUC-ROC & AUC-PR  \\  \hline
		1	&Curated 			& 75.03\% & 89.62\%  \\  
		2	&Unweighted 		& 78.23\% & 90.89\% \\   
		3	&Weighted-nClicks 	& 78.38\%  & 91.03\% \\ 
		\textbf{4}&\textbf{Weighted-CTR} & \textbf{78.61\%} & \textbf{91.22\%} \\  \hline
	\end{tabular}
	\caption{Offline evaluation of different CLSM models for sponsored search on human-labeled set.}	
	\label{tab:TravelResults}	
	\vspace{-8mm}
\end{table}

\subsubsection{Results}
As shown in Table \ref{tab:TravelResults}, the Unweighted model shows 1.27\% higher AUC-PR and 3.2\% higher AUC-ROC than the curated model on human-labeled evaluation set. Weighing further improves the unweighted model, with the best performing weighted model (Weighted-CTR) having 0.33\% higher AUC-PR and 0.38\% higher AUC-ROC. This demonstrates that on offline human labeled data, the weighted model performed equally well as the unweighted model in fact slightly improving the overall AUC numbers. These improvements were observed by running multiple iterations with different random weight initialization

\begin{table}
	
	\begin{tabular}{  c  c  c  c }		
		\hline
		Query Type & Model & AUC-ROC & AUC-PR \\  \hline
		& Curated & 79.07\% & 92.61\% \\ 
		Torso & Unweighted & 81.62\% &93.55\% \\ 
		& \textbf{Weighted-CTR} & \textbf{81.83\%} & \textbf{93.57\%} \\ \hline
		& Curated & 71.48\% & 85.84\% \\ 
		Tail & Unweighted & 75.45\% &87.74\% \\ 
		& \textbf{Weighted-CTR} & \textbf{75.85\%} & \textbf{88.40\%} \\ \hline
	\end{tabular}
	\caption{Evaluation of CLSM model for sponsored search on human labeled data for torso and tail queries.}	
	\label{tab:DecileResult}
		\vspace{-10mm}	
\end{table}

Table \ref{tab:DecileResult} shows the AUC gains when we break down the total gains into the torso and tail queries. We don't consider head queries here because head queries generally trigger exact matches rather than advanced matches. The weighted model shows much better AUC, especially in the tail bucket, with the weighted model having 0.66\% higher AUC-PR than unweighted model and 2.56\% higher AUC-PR than the curated model.


\begin{table*}	
	\begin{tabular}{  l  l  }		
		\hline
		Purchased product & Title of top - 1 returned product \\ \hline
		comfort control harness \textbf{large} & \textit{Weighted:} comfort control harness \textbf{xxl} blue \\
		&\textit{Unweighted:} comfort control harness \\ \hline
		\textbf{ford gt} red remote control car rc cars & \textit{Weighted:} licensed \textbf{shelby mustang} gt500 super \textbf{snake} rtr remote control rc cars \\
		&\textit{Unweighted:} lamborghini gallardo superleggera radio remote control car color yellow \\ \hline
		\textbf{nasa mars} curiosity rover \textbf{spacecraft} poster 13x19 &\textit{Weighted:} space shuttle \textbf{blastoff} poster 24x36 \\		
		&\textit{Unweighted:} imagination nebula motivational photography poster print 24x36 inch \\ \hline
		wonderful \textbf{wonder world clockmaker} & \textit{Weighted:} \textbf{alice country clover} ace hearts \\
		&\textit{Unweighted:} olympos \\ \hline	
		
	\end{tabular}	
	\caption{The table shows few examples of the top product returned by CLSM model trained on amazon co-purchased dataset. Weighted model is able to capture the specific intent, while products predicted by the unweighted model are more generic. Bold words are those non-overlapping words which contribute to the ten most active neurons at the max-pooling layer.}
	\label{tab:AmazonExample}	
	\vspace{-7mm}
\end{table*}


\subsection{Online Experiments on Sponsored Search}

\subsubsection{Dataset and Evaluation Methodology} Since the domain of sponsored search is very dynamic and an ad being relevant doesn't imply clickability, hence offline evaluation has limited power and is not sufficient. We perform online A/B testing of curated, unweighted and weighted-ctr model on Bing live traffic and report user interaction metrics.

\begin{table}	
	\begin{tabular}{ c  c  c  c  c  }		
		\hline
		  &   &\multicolumn{3}{c}{\% change of metrics in A}  	\\ 
  		  &   &\multicolumn{3}{c}{as compared to B}  	\\ \hline
		A 			& B 		& Clicks 	& Click-through & Bounce  	\\ 
		 			& 	 		& 		 	& rate (CTR)	& rate  	\\ \hline
		Unweighted	&Curated	&+10.11\%	&+10.08\%		&+2.85\%	\\ 
		\textbf{Weighted}	&Curated	&+22.78\%	&+23.13\%		&-5.63\%	\\ 
		\textbf{Weighted}	&Unweighted	&+11.51\%	&+11.85\%		&-8.25\%	\\ \hline		
	\end{tabular}	
	\caption{Online A/B testing of different CLSM models. Weighted model shows 23\% higher CTR and 5.6\% lower bounce rate as compared to curated model, while 11.8\% higher CTR and 8.2\% lower bounce rate as compared to unweighted model.}
	\label{tab:OnlineCompare}	
	\vspace{-8mm}
\end{table}

\subsubsection{Model Comparisons}
We assigned equal online traffic to each of the three models. We also ensured that every other setting on our ad stack is kept the same for these models. We compare these models on three major performance indicators: total clicks, CTR (Click through rate) and bounce rate.

\subsubsection{Results}
Table \ref{tab:OnlineCompare} shows A/B testing results, when different CLSM models are used for Ad Retrieval. First, as compared to the curated model, the unweighted model generated 10.1\% more clicks at 10.08\% higher click-through rate due to better exploration for tail queries, but bounce rate deteriorated by 2.8\%. Second, the Weighted model performed better than the curated model in all metrics. The weighted model generated 22.7\% more clicks at 23.1\% higher CTR while reducing bounce rate by 5.6\%. Third, the Weighted model also showed significantly better metrics than the unweighted model. Ads retrieved by weighted model generated 11.5\% more clicks while increasing CTR by 11.8\% and simultaneously reducing bounce rate by 8.2\%.
 
These results clearly demonstrate the impact of weighing on semantic model training. It is important to note here that apart from increasing the clicks and CTR, there is a huge reduction in terms of bounce rate when we move from unweighted to weighted. We see an increase in bounce rate when we move from curated to unweighted and we attribute this to numerous noisy pairs being directly added to training dataset with equal importance.


\subsection{Amazon Product Recommendation}
\subsubsection{Dataset and Evaluation Methodology}
In order to test the efficacy of weighing, independent of domain, we experiment with Amazon co-purchased dataset \cite{AmazonDataset} of related products. This dataset contains list of product pairs ($P_1$ , $P_2$) purchased together by users. It contains co-purchased data for 24 categories with a total of ~9 million unique products. We use this dataset to learn semantic models, which can be used for product recommendation task. On average, each product is associated with 36 co-purchased products. We make an 80-20 split of the dataset into train and evaluation set such that there is no product overlap between them. Models are evaluated on 20\% holdout set.
\begin{table}	
	\begin{tabular}{  c  c  c   c  c }		
		\hline
		Model & NDCG & NDCG & NDCG &  NDCG  \\
		& @1 & @3  & @5 &@10  \\ \hline
		Unweighted&63.71&69.08&72.84&79.75 \\
		\textbf{Weighted (Jaccard)}&\textbf{63.98}&\textbf{69.33}&\textbf{73.07}&\textbf{79.89} \\ \hline
	\end{tabular}	
	\caption{Evaluation of CLSM model trained for product recommendation task on amazon co-purchased dataset. NDCG metrics are reported on holdout set.}
	\label{tab:nDCGAmazon}
		\vspace{-10mm}	
\end{table}

\subsubsection{Model Comparison}
First, the unweighted CLSM model is trained by assigning equal weight to each co-purchased product pair. Second, weighted CLSM model is trained by assigning Jaccard index based weight to each product pair ($P_i$ , $P_j$). Jaccard index measures the similarity between two sets and is defined as the size of intersection divided by the size of the union. Let Nr($P_i$) represent set of neighbors of product $P_i$ i.e. set of products which were co-purchased with $P_i$, then Jaccard index between two products $P_i$ and $P_j$ can be defined as:
\begin{equation}
\label{eq:Jaccard}
	JI(P_i ,P_j) = \frac{\mid Nr(P_i) \cap Nr(P_j) \mid}  {\mid Nr(P_i) \cup Nr(P_j) \mid}
\end{equation}
The performance of models has been measured by mean Normalized Discounted Cumulative Gain (NDCG) \cite{nDCG}, and we report NDCG scores at truncation levels 1, 3, 5 and 10. 
\subsubsection{Results}

Table \ref{tab:nDCGAmazon} represents the efficacy of weighted model over unweighted in terms on NDCG scores. We see a gain of  0.27 in NDCG@1 and similar gains for other truncation levels. This result shows that weighing if domain agnostic.
\vspace{-2mm}	

\section{Discussion}
Each section of the table \ref{tab:AmazonExample} contains the title of purchased product and title of top-1 product returned by weighted and unweighted model. We further highlight those non-overlapping words between two titles that contribute to the ten most active neurons at the max-pooling layer. 

\begin{figure}[h]
	\centering
	\includegraphics[width=.48\textwidth]{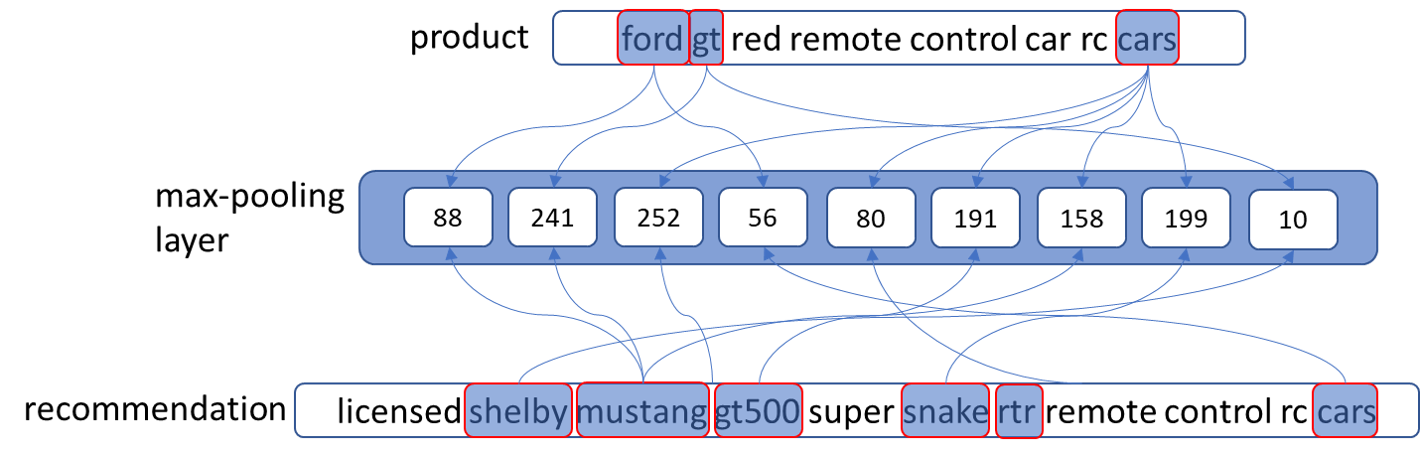}

	(a)	Weighted model
	\bigbreak
	\includegraphics[width=.48\textwidth]{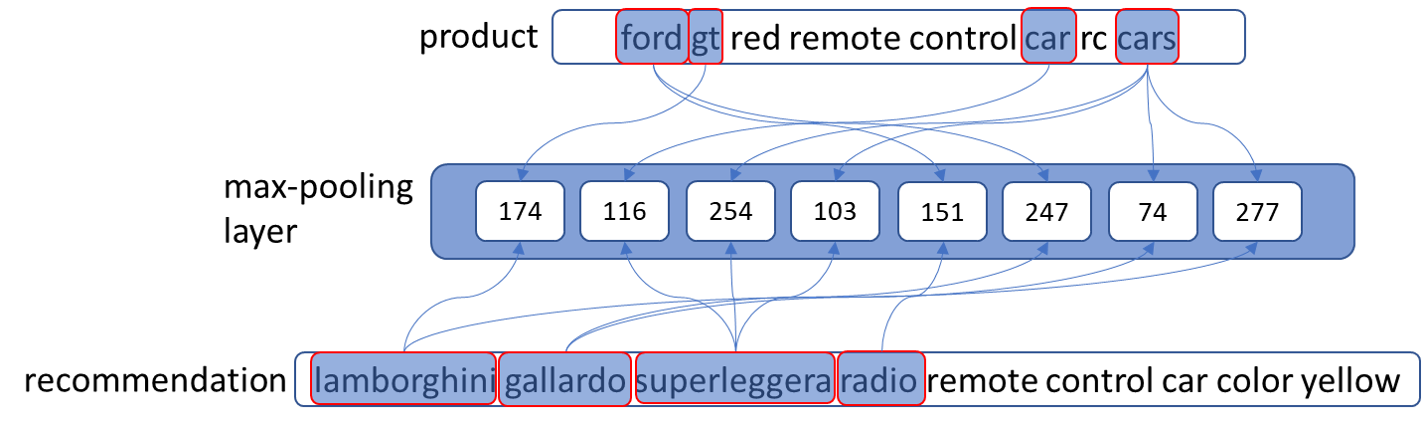}
	(b) Unweighted model
	\caption{Top neuron triggers at max-pool layer being mapped to contributing words. Weighted model maps top words like \textit{ford, gt, cars} to a particular car model by \textit{ford}. Whereas Unweighted model maps \textit{ford, gt, cars} to a car model from other company.} 
	\label{fig:example}
	\vspace{-6mm}
\end{figure}
In the first example, we see that apart from syntactic word matches, the weighted model is able to match the size as \textit{large} and \textit{xxl} in respective product titles appear among the ten most active neurons at the max-pooling layer. In the second example,  most active neurons for weighted model correspond to words like \textit{ford, gt; shelby, mustang and snake}, all of which refers to a particular car model by \textit{ford} named \textit{"ford mustang shelby GT500 super snake"}. Whereas for the unweighted model, most active neurons correspond to words like \textit{cars, car, remote, rc, ford; lamborghini, gallardo, superleggera, remote}. So unweighted model only captures the general intent, rather than specific intent captured by the weighted model.
To further examine the learning, we trace the neurons with high activation at the max-pooling layer to the words from product title. 
Figure \ref{fig:example} shows that while the weighted model's retrieval is governed by the similarity occurring between words related to a particular car, the retrieval of the unweighted model is governed by a general similarity between two different model of cars. We see that due to weighing, we are able to retrieve the specific car as opposed to any generic car. Similar observations were made in example 3 and 4 where weighted model recommended more specific products as opposed to a generic poster or a book.
\vspace{-2mm}

\section{Conclusion}
This paper developed the cost-sensitive approach to training semantic models for IR. It extended the pair-wise loss of CLSM model by re-weighing the train data points and evaluated various weighing techniques for the same. A/B testing of proposed model on Bing sponsored search showed significant improvements in click-through rates and bounce rates. The proposed idea is general and is applicable to different tasks like retrieval, recommendation, classification etc.

\clearpage

\bibliographystyle{ACM-Reference-Format}
\bibliography{Paper_Bibliography}

\end{document}